\documentclass[]{article}

\usepackage{PRIMEarxiv}

\usepackage[T1]{fontenc}    
\usepackage{hyperref} 
\usepackage{url}            
\usepackage{booktabs}       
\usepackage{amsfonts}       
\usepackage{nicefrac}       
\usepackage{microtype}      
\usepackage{lipsum}
\usepackage{fancyhdr}       
\usepackage{graphicx}       
\usepackage{cleveref}
\usepackage{wrapfig}

\usepackage[]{todonotes}

\graphicspath{{images/}}     

\title{Proteome-scale Deployment of Protein Structure Prediction Workflows on the Summit Supercomputer
\thanks{Notice: This manuscript has been authored in part by UT-Battelle, LLC under Contract No. DE-AC05-00OR22725 with the U.S. Department of Energy. The United States Government retains and the publisher, by accepting the article for publication, acknowledges that the United States Government retains a non-exclusive, paid-up, irrevocable, world-wide license to publish or reproduce the published form of this manuscript, or allow others to do so, for United States Government purposes. The Department of Energy will provide public access to these results of federally sponsored research in accordance with the DOE Public Access Plan (http://energy.gov/downloads/doe-public-access-plan).
}}

\author{
  Mu Gao \\
  Center for the Study of Systems Biology \\
  Georgia Institute of Technology \\
  Atlanta, Georgia, USA\\
  \texttt{mu.gao@gatech.edu} \\
   \And
  Mark Coletti \\
  Oak Ridge National Laboratory \\
  Oak Ridge, Tennessee, United States \\
  \texttt{colettima@ornl.gov} \\
  \And
  Russell B. Davidson \\
  Oak Ridge National Laboratory \\
  Oak Ridge, Tennessee, United States \\
  \texttt{davidsonrb@ornl.gov} \\
   \AND
   Ryan Prout \\
 Oak Ridge National Laboratory \\
  Oak Ridge, Tennessee, United States \\
   \texttt{proutrc@ornl.gov} \\
   \And
   Subil Abraham \\
Oak Ridge National Laboratory \\
  Oak Ridge, Tennessee, United States \\
   \texttt{abrahams@ornl.gov} \\
   \And
   Benjam\'in Hern\'andez \\
   Oak Ridge National Laboratory \\
   Oak Ridge, Tennessee, United States \\
   \texttt{hernandezarb@ornl.gov} \\
   \And
   Ada Sedova \\
  Oak Ridge National Laboratory \\
  Oak Ridge, Tennessee, United States \\
   \texttt{sedovaaa@ornl.gov} \\
}

\begin{document}
\maketitle

\begin{abstract}
Deep learning has contributed to major advances in the prediction of protein structure from sequence, a fundamental problem in structural bioinformatics. With predictions now approaching the accuracy of crystallographic resolution in some cases, and with accelerators like GPUs and TPUs making inference using large models rapid, fast genome-level structure prediction becomes an obvious aim. Leadership-class computing resources can be used to perform genome-scale protein structure prediction using state-of-the-art deep learning models, providing a wealth of new data for systems biology applications. Here we describe our efforts to efficiently deploy the AlphaFold2 program, for full-proteome structure prediction, at scale on the Oak Ridge Leadership Computing Facility's resources, including the Summit supercomputer. We performed inference to produce the predicted structures for 35,634 protein sequences, corresponding to three prokaryotic proteomes and one plant proteome, using under 4,000 total Summit node hours, equivalent to using the majority of the supercomputer for one hour. We also designed an optimized structure refinement that reduced the time for the relaxation stage of the AlphaFold pipeline by over 10X for longer sequences. We demonstrate the types of analyses that can be performed on proteome-scale collections of sequences, including a search for novel quaternary structures and implications for functional annotation.
\end{abstract}


\section{Introduction}
\label{sec:intro}
Understanding the function of all of the proteins coded for by an organism's genome---its proteome---is a grand challenge for the biological sciences, with implications for synthetic biology, medicine, molecular genetics, and many other efforts \cite{gao2021high}. Knowledge of the atomic structure of a protein can help with multiple aspects of biological research \cite{sillitoe2014cath}. For instance, it can inform our understanding of protein function, such as small-molecule binding pockets that one can only make sense of in three-dimensional space. A structure can enable comparisons across databases \cite{holm1993dali}, analyses such as ligand-binding \cite{gao2013apoc} or protein-protein interface determination \cite{gao2010ialign}, and molecular docking and simulation \cite{legrand2020gpu,vermaas2020supercomputing}. However, obtaining protein structures experimentally is time consuming, costly, and difficult; for instance, only about 17\% of the human proteome has an experimentally determined three-dimensional structure \cite{jumper2021highly}. Therefore, being able to accurately predict protein structure computationally has been a goal for decades. Over the past several years, exciting breakthroughs have been made in the use of deep learning together with advanced structural informatics methods to improve protein structure prediction \cite{seni2020improved,xu2019distance,gao2019destini}. Most recently, Google DeepMind's AlphaFold2 achieved unprecedented accuracy in the Critical Assessment of protein Structure Prediction experiments (CASP14) \cite{jumper2021highly}. The accuracy of this method is attributable partially to the size of the neural network as well as to innovations in the network architecture. The use of accelerators such as tensor processing units (TPUs) or graphics processing units (GPUs) to both train and perform inference is important for high performance, and essential for genome-scale applications. With this breakthrough comes the possibility of generating large datasets of high-accuracy predictions of protein structures containing orders of magnitude more elements than existing crystallographic databases hold. Thus, deployment of AlphaFold at scale to predict the structure of an organism's full proteome is a natural desire. 

Considerable advances in next-generation sequencing technologies have driven an exponential growth in the number of sequenced genomes \cite{tatusova2016ncbi,uniprot2018uniprot}. Computational approaches that can take advantage of this massive amount of data and provide important information about the proteome will help advance biology into a new digital and big-data age. In fact, some of the leading names in high-performance computing (HPC) have recognized not only the need for HPC in biology, but also the potential in this area, calling this next phase of computational biology ``the digital biology revolution." \footnote{https://blogs.nvidia.com/blog/2021/07/07/ceo-unveils-cambridge-1/} However, fewer computational biology applications have made use of HPC or even code acceleration with GPUs than in other fields such as astrophysics. Barriers have included heterogeneous sizes and file formats of biological datasets and the multiple steps required for processing and analysis, each of which may use specialized programs often developed by third-parties. These factors are in part a result of variability in genome structure across taxa, and the multi-scale and multi-disciplinary complexity of gene expression, regulation and function.    
The convergence of HPC and artificial intelligence (AI), especially deep learning (DL), promises to open the door to the use of HPC resources for large-scale biological problems. These tools are designed for high-level developers, and can efficiently incorporate highly non-linear and multi-dimensional information to provide accurate predictions; many DL programs can also automatically make use of both GPUs and parallel processors. As such, use of DL and other AI methods in the biosciences has recently flourished \cite{ramsundar2019deep}. The success of DL to provide accurate prediction enabled by the use of high-performance hardware such as GPUs is evidenced by AlphaFold2. Here we describe the implementation and deployment of a proteome-scale, high-throughput protein structure prediction workflow using AlphaFold2 on the Summit supercomputer at the Oak Ridge Leadership Computing Facility (OLCF). We detail the performance optimizations and workflow management aspects of this deployment, and demonstrate two useful applications of large datasets of predicted structures: for informing functional annotation, and for discovering new assemblies of protein domains.

\section{Background and Related Work}
\label{sec:background}
To our knowledge, the only previous deployment of AlphaFold2 at proteome-scale was performed by the Google DeepMind AlphaFold team in collaboration with the European Molecular Biology Laboratory's (EMBL) European Bioinformatics Institute (EBI). This dataset has been deposited for public use and contains 360,000 predicted structures across 21 model-organism proteomes; the goal of the project is to predict the structures of all proteins in the UniRef90 sequence database \cite{alphafold2021DB}, a database that can be considered a (growing) list of all known non-redundant protein sequences that is continuously updated with protein representatives from clusters with 90\% sequence identity \cite{uniprot2018uniprot}. Currently, it contains over 100 million sequences, and will require close to 280 times more compute time than was used to generate the initial 360,000 entries in the AlphaFold Protein Structure Database. It is therefore important for the scientific community that additional resources be enlisted to help reduce the time it takes to create this dataset: such a compendium of information could lead to fundamental breakthroughs in our understanding of both protein structure and function \cite{SERPELL2021167231}.

The deployment of high-throughput computational biology workflows on HPC systems has become more common. Even for large sets of mostly parallel calculations, HPC systems have been used \cite{joubert2018parallel,acharya2020supercomputer,glaser2021high}. Often providing larger numbers of compute nodes simultaneously, HPC resources can enable a large high-throughput job to be performed in a fraction of the total time than is possible in the cloud, helping to address urgent and timely problems \cite{vermaas2020supercomputing}. Furthermore, shared parallel filesystems allow for rapid, in-place data analysis on large output datasets using parallel tools \cite{glaser2021high}. The ability to make use of either the high-speed interconnect for traditional parallelization with MPI \cite{joubert2018parallel}, or alternately, a dataflow execution model \cite{ossyra2019porting,ossyra2019highly}, or a combination of the two \cite{acharya2020supercomputer}, helps with design of flexible, multi-task workflows that are often found in computational biology.

\subsection{Leadership-scale HPC Workflows}
While numerous workflow management tools exist, few have been deployed at scale on the largest supercomputers. The FireWorks workflow management software \cite{jain2015fireworks} is frequently used for materials science workflows on the National Energy Research Scientific Computing Center (NERSC) resources; it was first tested at the scale of Summit for high-throughput molecular docking screens \cite{glaser2021high,vermaas2020supercomputing}. The Dask library was designed to enable parallel computing in Python by distributing scientific calculations such as linear algebra operations and singular-value decomposition \cite{rocklin2015dask,dask}; it can also be used purely as a distributed workflow manager via Python \texttt{subprocess} calls. On Summit, Dask for the former use case was tested on small numbers of nodes  \cite{hernandez2020performance}, and deployed on larger numbers of nodes for parallel, distributed dataframe and database query processing tasks \cite{glaser2021high}; it was deployed as a workflow manager for large-scale evolutionary algorithms at the scale of 500 Summit nodes \cite{gremlin2021,coletti2019troubleshooting}.

\section{Methodology}
\label{sec:methodology}
Here we describe our optimized workflow for running the AlphaFold2 pipeline in a high-throughput HPC setting, and testing on the OLCF resources. 
For this pipeline, pre-processing steps and feature generation were performed on the Andes analysis cluster. The Summit supercomputer with its GPUs was used in large parallel batches to run the DL inference as well as the geometry optimization calculations on the final models. These last two steps utilized Dask as the workflow management software. 

The Summit supercomputer is an IBM system containing approximately 4,600 IBM Power System AC922 compute nodes. Each node contains two IBM POWER9 processors and 6 NVIDIA Tesla V100 accelerators. Each processor is connected via dual NVLINK connections capable of a 25GB/s transfer rate in each direction\footnote{\url{https://docs.olcf.ornl.gov/systems/summit_user_guide.html}}. Andes is a commodity-type Linux cluster with 704 nodes, each containing two 16-core 3.0 GHz AMD EPYC 7302 processors with AMD’s Simultaneous Multithreading (SMT) Technology and 256GB of main memory\footnote{\url{https://docs.olcf.ornl.gov/systems/andes_user_guide.html}}. Some preliminary testing of the pipeline was also performed on the Partnership for an Advanced Computing Environment (PACE) resources at Georgia Institute of Technology. PACE houses the Phoenix cluster, a mixture of CPU and GPU nodes, with approximately 1100 CPU nodes and 100 GPU nodes. GPU nodes consist of dual Intel Xeon Gold 6226 CPUs each with 12 cores, and 4 x RTX6000 GPUs (24GB memory), totaling 96GB GPU memory and 24 CPU cores per node. 

\subsection{AlphaFold2 on the Summit Supercomputer using Singularity}
The computing power of the Summit supercomputer is an important resource for proteome-scale protein structure prediction; one of our aims is to supplement the AlphaFold/EMBL-EBI database with organisms of interest to the Department of Energy (some of which may not yet be found in the UniRef90 database, the target of AlphaFold/EMBL-EBI efforts), and to produce full-genome structural predictions for organisms central to our research. Challenges to deploying AlphaFold on Summit include the Power9 CPU architecture that prevents the use of package managers to download pre-compiled binaries, combined with the use of build systems non-standard in the HPC setting. AlphaFold provides a Docker\footnote{\url{https://www.docker.com/}} recipe for building all of its dependencies and packaging, but that recipe is not directly usable on Summit or on other systems which do not support Docker. Summit does not yet support container builds, but it does provide a functioning container runtime with Singularity. As described in \cite{gao2021high}, we used an external POWER9/NVIDIA V100 GPU system to build a Singularity\footnote{\url{https://sylabs.io/singularity/}} container that could run on Summit using Podman\cite{gantikow2020rootless} as the container build mechanism. We altered the AlphaFold Dockerfile to build JAX and other components from source, and to remove feature generation steps which did not require GPUs and thus could be pre-computed on Andes. The final image was converted to Singularity's format for running on Summit with Singularity.

\subsection{Optimization of the AlphaFold Method for the High-throughput Setting}
The release of the source code and neural network models of AlphaFold (version 2.0.1 and above) by DeepMind sparked the interests of both the biological and computational research communities \cite{skolnick2021alphafold}. However, the code released by AlphaFold was designed for small-scale studies, and not optimized for large-scale deployments, especially on an HPC system designed for many users such as at the OLCF. This is in part due to the use of three task stages each with very different computing requirements and characteristics described below. 

\subsubsection{Input feature generation on CPU resources}
Given an input protein sequence, AlphaFold first runs multiple searches for homologous sequences and generates the input multiple sequence alignment (MSA), using third-party sequence alignment tools such as those provided in the HMMER \cite{eddy2011accel} and HH-suite \cite{steinegger2019hh} packages. These programs are run against several sequence libraries (UniProt \cite{uniprot2018uniprot}, MGnify \cite{mitchell2019mgnify}, BFD \cite{steinegger2019protein}), and sequences of experimentally solved structures collected from the Protein Data Bank \cite{berman2000pdb}. The total storage amount of these sequence libraries is about 2.1 TB for the full dataset \cite{jumper2021highly}, and 420 GB for a reduced dataset described below. This first computational task, while I/O and memory intensive, uses CPU-based codes that make extensive use of vectorization as their primary performance strategy; tasks are run in batches and alignment of a single sequence to tens of thousands of sequences in the databases is finished in minutes. On the OLCF resources, the Summit computer is the most expensive and precious resource in terms of cost of maintenance, hardware, and power, and also in terms of the competitiveness of compute allocations. Therefore, it would be wasteful to use Summit time to deploy the original AlphaFold pipeline, wherein GPUs would be idle for the first part of the calculation. A recent version of AlphaFold was designed to allow for parallel execution of the CPU portion and the GPU portion of the pipeline when running in larger batches \cite{zhong2022parafold}; however, our tests showed that the compute patterns, resource requirements, and execution times are too imbalanced to make this the optimal solution when running inference at scale on Summit. We therefore pre-computed input features on Andes.

For HHSuite, the number of file reading tasks performed by one alignment to the database can be large, and this becomes the bottleneck on shared filesystems due to traffic on the metadata servers and disk accesses\footnote{\url{https://github.com/soedinglab/hh-suite/wiki\#running-hhblits-efficiently-on-a-computer-cluster}}. Unfortunately, due to the shared-use design of open-science HPC resources such as the OLCF, it is not possible to copy these large databases into compute node memory or onto NVME burst buffers and leave them there for multiple jobs. Therefore, the time saved from using this type of memory can be cancelled-out by repeated copying with every job allocation. In order to reduce I/O overhead, we created 24 identical copies of the reduced sequence libraries on the parallel filesystem using \texttt{mpiFileUtils}\footnote{\url{https://hpc.github.io/mpifileutils/}}, and ran 4 parallel jobs on each copy of the library. After testing with the full 2.1 TB dataset, we moved to using a reduced version, which is obtained by removing identical and near-identical sequences in the largest of the sub-datasets, the BFD. According to DeepMind's benchmark study, it yields virtually identical performance as the results from the full dataset \cite{tuny2021human}. The sequence searches yield MSAs between the input target sequence and its homologous sequences found in the libraries, and also structures of homologs to be used as templates from aligning to the PDB databases. AlphaFold then builds input features from the MSAs and the structural templates. The structural features are only used by two of the five DL models that each output a structure in AlphaFold2. The other three models take only the sequence features to make predictions. Thus, the most important features are the MSAs, which dictate the final quality of all predicted structures. The input features are then fed into the DL neural network models that eventually predict the three-dimensional coordinates of the input sequence. 

\subsubsection{Inference}
\label{sub:infmethod}
The model inference procedure is iterated in multiple recycles, feeding back a predicted 3D structural model out of each inference cycle and regenerating input features as feedback. In practice, multiple iterations are essential and the official release of AlphaFold uses three recycles. Overall, the calculations in this stage involve many tensor operations, thus a newer GPU, such as those available on Summit, can dramatically accelerate the calculation speed. The official release of AlphaFold version 2.0.1 provides two presets, \texttt{reduced\_dbs} and \texttt{casp14}, that combine a few key configuration parameters for application or benchmark purpose. The \texttt{casp14} preset was adopted during the CASP14 competition; it employs eight ensembles reulting in approximately eight times the computational cost in comparison to the single ensemble preset defined in \texttt{reduced\_dbs}. Both presets use a fixed number of 3 recycles. The \texttt{reduced\_dbs} is the preset that DeepMind used in their proteome-scale applications of AlphaFold \cite{tuny2021human}. While this preset is efficient and often sufficient for most protein sequences, it is also known that some challenging protein sequences may not be folded properly in a short number of recycles \cite{jumper2021highly}. To address these challenging sequences without dramatically increasing run time, we implemented a method to dynamically control recycle numbers first proposed in ColabFold \cite{mirdita2021colabfold}. In this strategy, the change of the protein residue contact distogram is calculated after each recycle in comparison to the previous recycle. If the change is less than a threshold value, it is consider converged and the program will stop recycling. In our implementation, we use two threshold cutoff values of 0.5 and 0.1 for our customized \texttt{genome} and more stringent \texttt{super} presets, respectively. Additionally, we increased the maximum number of recycle values to 20, but also reduced it progressively (to a minimum of 6) with increased input sequence length, if the length is longer than 500 amino acid residues (AAs). While our implementation also provides other options that are useful to reduce memory requirements for very long sequences (e.g. $>$ 2500 AAs), we do not report predicted structures for these large proteins here as they comprise less than 1\% of the sequences. Such long sequences are best processed separately with adjusted parameters to more appropriately balance accuracy and time to solution.

\subsubsection{Geometry optimization}
\label{sub:geom}
The model inference generates high-quality 3D structural models that are typically sufficient as-is for most research purposes. However, it is not uncommon that these models still contain structural flaws that would not be found in experimental crystal structures and may be considered non-physical. These deviations in geometry have been classified for the CASP assessments as  ``clashes" and ``bumps," with the following definitions:  a clash is a C$\alpha$-C$\alpha$ pairwise distance $<$ 1.9 \AA, a bump is a C$\alpha$-C$\alpha$ pairwise distance $<$ 3.6 \AA. Models are considered to be ``clashed" if they contain more than 4 clashes or 50 bumps \cite{tress2005assessment}. For deposition of model structures into databases, it may be considered good practice to remove clashed structures for aesthetic purposes. To address this issue, the final step of AlphaFold is to remove these clashes using a standard molecular mechanics optimization procedure called the ``relaxation." This consists of an optimization of the Cartesian coordinate geometry using a minimization with respect to some function, often the potential energy as defined by a molecular mechanics Hamiltonian and force field. The critical task is to maintain the accurate structure provided by the inferred model while removing the non-physical clashes and bumps; only small perturbations to the overall structure are desired. AlphaFold employs the OpenMM program \cite{eastman2017openmm} for this procedure, which can run either purely on CPUs or take advantage of GPUs. In the original AlphaFold program, the CPU version is used. We have found that the GPU version can provide exceptional performance on Summit \cite{ossyra2019porting}, so we designed a relaxation protocol that used OpenMM's GPU platform. We also decoupled this step into its own separate workflow. 

For each predicted structure, atoms were assigned force field parameters and hydrogen atoms were added. Then, a single energy-minimization calculation was performed with an unlimited number of optimization steps until the energy difference between steps reached a convergence criteria (2.39 kcal $\cdot$ mol$^{-1}$). The force field and minimization protocol mirror those used in AlphaFold2 exactly, including the application of a harmonic restraint to all non-hydrogen atoms using a force constant of 10 kcal $\cdot$ mol$^{-1}$ $\cdot$ \AA$^{-2}$. Where our method does differ from AlphaFold2 is in the number of energy minimization calculations that may be attempted on each structural model. The original AlphaFold2 procedure checks for ``violations'' (e.g. clashes and bumps); if any violations are found, another iteration of minimization calculations are performed on the model. However, at this point in the pipeline, the model structure is described with a molecular mechanics force field that strongly destabilizes non-physical interactions between any atoms in the model (beyond those defined by C$\alpha$-C$\alpha$ distances). More than a single energy minimization calculation is rarely needed, so we removed the unnecessary violation calculations and the possibility for repeated energy minimization calculations. We compared both the runtime and performance of the geometry optimization between our protocol and the original AlphaFold2 version to test ability of our modified method to remove violations as well as to speedup the time-to-solution. For gauging the ability of our method to perform adequate relaxation, we used proteins from the CASP14 assessment for which crystal structures are available for comparison. Relaxation with the original AlphaFold method was performed on the PACE Phoenix cluster. Our protocol using the CPU version of OpenMM was tested on Andes using a full node (2 AMD EPYC 7302 Processors with 16 cores each) with the default threading scheme in OpenMM that sets the thread count based on the number of cores it detects. Our protocol using Summit assigned a single CPU core and a single GPU for each minimization calculation; this is more than sufficient for small systems of thousands to tens of thousands of atoms. Therefore, 6 instances of our geometry optimization task can be run on one Summit node for the full-scale workflow.

\subsection{Deployment of Inference Workflow on Summit with Dask}
\label{sub:workflow}
After pre-calculating the input features, the DL inference is deployed using a workflow manager with a dataflow execution model where workers receive tasks as soon as they are free to accept them from a queue. AlphaFold uses five different models to perform inference on each target protein sequence, producing five different predicted structures. Therefore, computing tasks are composed of pairs of DL models and target sequences. In each task, we run one DL model given the input features of one target sequence, and it generates one structural model. This task decomposition strategy helps with load distribution and balance in a large-scale parallel-job run.

\begin{figure}[htb]
    \centering
    \includegraphics[width=0.6\linewidth]{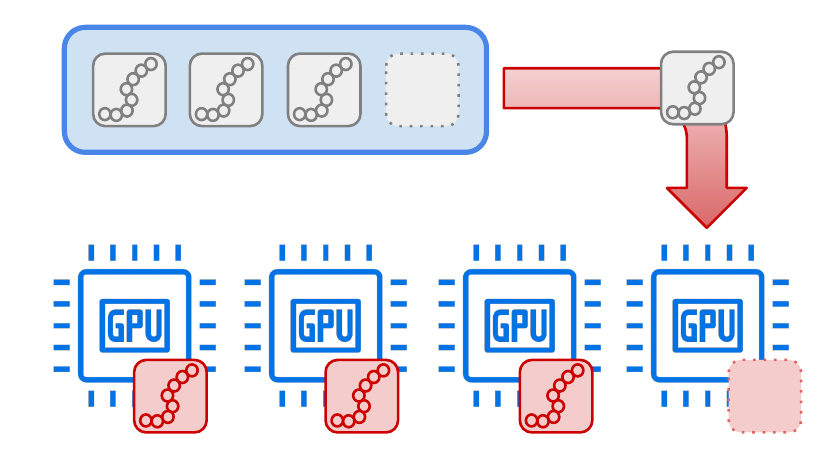}
    \caption{Illustration of the Summit AlphaFold workflow for asynchronously processing a batch of inferences from input features. The top oval box represents the Dask scheduler task queue that has three protein inputs waiting for an available worker. The four bottom GPU icons represent Dask workers with an associated GPU running the inference. The arrow shows the Dask scheduler assigning the next task in its queue to a Dask worker that recently finished processing a prediction. This process repeats until the scheduler queue is empty via a dataflow execution model.}
    \label{fig:workflow}
\end{figure}
Fig.~\ref{fig:workflow} shows the Summit AlphaFold2 workflow. Dask \cite{dask} is used as the workflow management tool, assigning input features for each of the five models for each protein target to the AlphaFold inference portion, using the following steps from a single Summit batch submission script (Summit uses IBM's LSF batch queueing system\footnote{\url{https://en.wikipedia.org/wiki/IBM_Spectrum_LSF}}):

\begin{enumerate}
    \item The Dask scheduler is started.
    \item Dask workers, one per GPU for all Summit nodes used, are started; they use a JSON file written by the Dask scheduler to register with the scheduler.
    \item The main driving Python script is started. \label{line:mainscript}
    \begin{enumerate}
        \item A Dask client connects to the scheduler.
        \item A text file is read that contains a list of all the protein targets to be processed.
        \item The list of targets are sorted in descending order of sequence length. \label{item:sort}
        \item All input tasks are added to the scheduler task queue via a single Dask \texttt{client.map()} call.
        \item As each Dask task is completed, statistics about that task, such as the start and end processing times, are appended to a CSV file.
    \end{enumerate}
\end{enumerate}

A concern with high-throughput biological workflows is load balance, due to the variable range in input sizes \cite{thavappir2021adressing}. We implemented a greedy approach to load balancing by sorting proteins in descending order by sequence length (\ref{item:sort}), allowing for lengthier processing to happen earlier in the run. Smaller tasks fill in gaps later. With a random task-processing order, some of longer-running tasks could happen at the end and be assigned to a single worker to run sequentially; in those scenarios, the run would continue until those tasks complete even though the remaining workers have finished all the remaining tasks and are idle. Some of the proteins are too large to fit onto the memory of a standard Summit node, and after a run these proteins will have failed to process. For these we used Summit's additional high memory nodes with 2TB of DDR4 memory, 192GB of High Bandwidth Memory (HBM2).\footnote{\url{https://docs.olcf.ornl.gov/systems/summit_user_guide.html}} The LSF script for processing AlphaFold inference on Summit used three \texttt{jsrun} statements; \texttt{jsrun} is IBM's version of \texttt{mpirun} for Summit. The first was to run a Dask scheduler using just two cores. The second \texttt{jsrun} allocated a Dask worker per GPU. The third \texttt{jsrun} dedicated a single core for the controlling Python script that was the Dask client; this was the script that read the file of proteins to be processed, assigned tasks via Dask for AlphaFold inference, and created a CSV file of the processing times for each task, as listed in \ref{line:mainscript} above.

\subsection{Deployment of Geometry Optimization Workflow on Summit with Dask}
\label{sub:relaxworkflow}
The geometry optimization portion of the pipeline was also deployed with a Dask workflow, in a nearly identical setup as described in~\ref{sub:workflow}. Instead of input features, inputs were the unrelaxed model structures output by the inference workflow. Instead of model inference, the relaxation steps described in~\ref{sub:geom} were run on each structure, using one GPU per task.

\section{Results and Discussion}
\label{sec:results}
Here we describe the results of the first proteome-scale deployments of our AlphaFold structure prediction workflow on the OLCF resources, including statistics on performance, confidence, and quality. The total number of structures predicted is five times the total number of input target sequences, due to the generation of a structure for each of 5 different models. The top model is chosen based on the confidence scores predicted by AlphaFold2. We used the output pTMS value for this choice. We predicted the structures of all proteins with a sequence length of less than 2500 AAs for the following species: prokaryotes \textit{Pseudodesulfovibrio mercurii}, \textit{Rhodospirillum rubrum}, and \textit{Desulfovibrio vulgaris} strain Hildenborough, and the plant species \textit{Sphagnum divinum} as a representative eukaryote. All four species are important to research efforts within the Department of Energy's Office of Biological and Environmental Research. \textit{P. mercurii} is a newly reclassified species of bacteria found to be extremely efficient at methylating mercury; methylated mercury is highly toxic and thus this organism is important to environmental science research \cite{gilmour2021pseudodesulfovibrio}. \textit{R. rubrum} is a photosynthetic bacterium that has been studied for its important metabolic pathways such as light harvesting, elemental sulfur production \cite{munk2011complete}, and recently, a newly discovered ability to produce ethylene, useful for bioproduction \cite{north2020nitrogenase}. \textit{D. vulgaris} Hildenborough is the first sulfate-reducing bacterium with a sequenced genome and has been chosen to be a model organism by the ENIGMA collaboration; these organisms play a key role in biogeochemical cycles on the planet, and may be useful for bioremediation, biomedical research, and biofuels \cite{wall2021deletion,heidelberg2004genome}. \textit{S. divinum} is a species of peat moss; peat is responsible for the sequestration of over half of the earth's carbon, and under warming temperatures, may release this carbon, adding to greenhouse gases. Formerly classified as the \textit{S. magellanicum} species, \textit{S. divinum} has recently been considered divergent enough to be classified as a new species, and may be able to better withstand varying climates \cite{hassel2018sphagnum}; thus it is important to investigate this species at genome-scale to aid remediation efforts \cite{malhotra2020peatland,kluber2020constraints}. The number of final (top) predicted protein structures for each species is 3446, 3849, 3205, and 25134 for \textit{P. mercurii}, \textit{R. rubrum}, \textit{D. vulgaris} Hildenborough, and \textit{S. divinum}, respectively.

\subsection{Feature Generation}
 Feature generation using the Andes CPU cluster required approximately half the resources to run than inference on Summit, when measured solely by node-hour. For one of the bacterial proteomes containing 3205 protein sequences with a mean of 328 AAs, feature generation took about 240 Andes node hours (vs.~about 400 Summit node hours for inference). 
 After testing, we found the reduced sequence dataset was sufficient for accuracy and better for large-scale applications because it dramatically reduced the storage and copy requirements, and the I/O overhead.
 
 \subsection{Model Inference}
 The performance and the run-time costs of the presets described in~\ref{sub:infmethod} are compared to the official AlphaFold presets in Table~\ref{tab:presets} using 559 protein sequences from \textit{D. vulgaris} Hildenborough as a benchmark. The length of the benchmark sequences ranges from 29 to 1266 AAs with a mean of 202 AAs. 
 All runs are conducted on 32 Summit nodes, except for the \texttt{casp14} run on 91 nodes. 
 Results of the eight longest sequences for the \texttt{casp14} runs are missing due to out-of-memory errors caused by high ensemble number. Wall time is total wall-clock time for all 559 input sequences and includes overhead, which is about 16\% of the total time in the \texttt{super} preset run. Means were calculated on the top-ranked model, ranked by either predicted LDDT (pLDDT) or predicted TM-score (pTMS). Higher pLDDT or pTMS score is better; their perfect values are 100 and 1.0, respectively. Best performance is indicated by bold font. The pLDDT metric measures local model quality, and provides a good measure for individual domains, i.e., single-domain protein folds. However, as most proteins have multiple domains, a better metric to evaluate the global model quality is the pTMS.
 
 The \texttt{genome} and \texttt{super} presets yield better top-ranked models in comparison to the \texttt{reduced\_dbs} or \texttt{casp14} presets, thanks to longer recycles. If we use a pLDDT score of 70 as the high-quality cutoff, the \texttt{genome} and \texttt{super} presets deliver a high-quality top model in 80\% of target sequences, whereas the \texttt{reduced\_dbs} delivers high-quality models in 77\% of cases. If we use pTMS of 0.60 as the threshold for high-quality global model, we find that the \texttt{genome} and \texttt{super} both provide a high-quality model for 62\% of targets, in comparison to 59\% of \texttt{reduced\_db}. Our presets have higher run time costs in comparison to the \texttt{reduced\_db}, but significantly better than the \texttt{casp14} preset. Since the performance of \texttt{genome} and \texttt{super} presets are very close, and the former costs slightly less, we adopt it as the main preset for our proteome-scale prediction workflows.
 
 \begin{table}[htbp]
 \begin{center}
  \caption{Benchmark tests of presets on 559 protein sequences$^{\mathrm{a}}$}
  \label{tab:presets}
  \begin{tabular}{ccccc}
    \hline
    \textbf{Preset}& \textbf{Mean pLDDT} &\textbf{Mean pTMS}  &\textbf{Count} &\textbf{Walltime$^{\mathrm{a}}$}\\
    \hline
    \textbf{reduced\_db}&78.4&0.631&559&\textbf{44}\\		
    \textbf{genome}&79.5&0.644&559&50\\
    \textbf{super}&\textbf{80.7}&\textbf{0.650}&559&58\\
    \textbf{casp14$^{\mathrm{c}}$}&78.6&0.631&551&$>$150\\
\hline

\multicolumn{5}{l}{$^{\mathrm{a}}$Means across top structure ranked by either pLDDT or pTMS. }\\
\multicolumn{5}{l}{$^{\mathrm{b}}$Wall time (min) includes overhead. }\\
\multicolumn{5}{l}{$^{\mathrm{c}}$8 longest sequences missing due to out-of-memory errors. }
\end{tabular}
\vspace{-0.5cm}
\end{center}
\end{table}
 
Although the mean improvement using either metric for \texttt{genome} or \texttt{super} presets appears small, it is important to note that the improvement is not uniformly distributed among individual sequences. If that were true, one may argue that the extra time spent may not be worth it. Rather, the improvement is largely attributed to a few cases where significantly better models emerge from longer recycles. For example, with the \texttt{super} preset run, about 45\% of the improvement measured by total sum of pTMS comes from 28 (5\%) targets with a significant 0.1 or higher improvement in their individual model pTMS, and 74\% of improvement comes from 68 (12\%) targets with an pTMS improvement above 0.05. Virtually all of these models are generated from numerous recycles (close to 20, the upper limit), with a  mean of 19. Therefore, this option is useful for challenging targets where a better chance of a good model may be possible with longer recycles.

\subsection{Inference Workflow}
\begin{figure}[t]
    \centering
    \vspace{-0.7cm}
    \includegraphics[width=\linewidth]{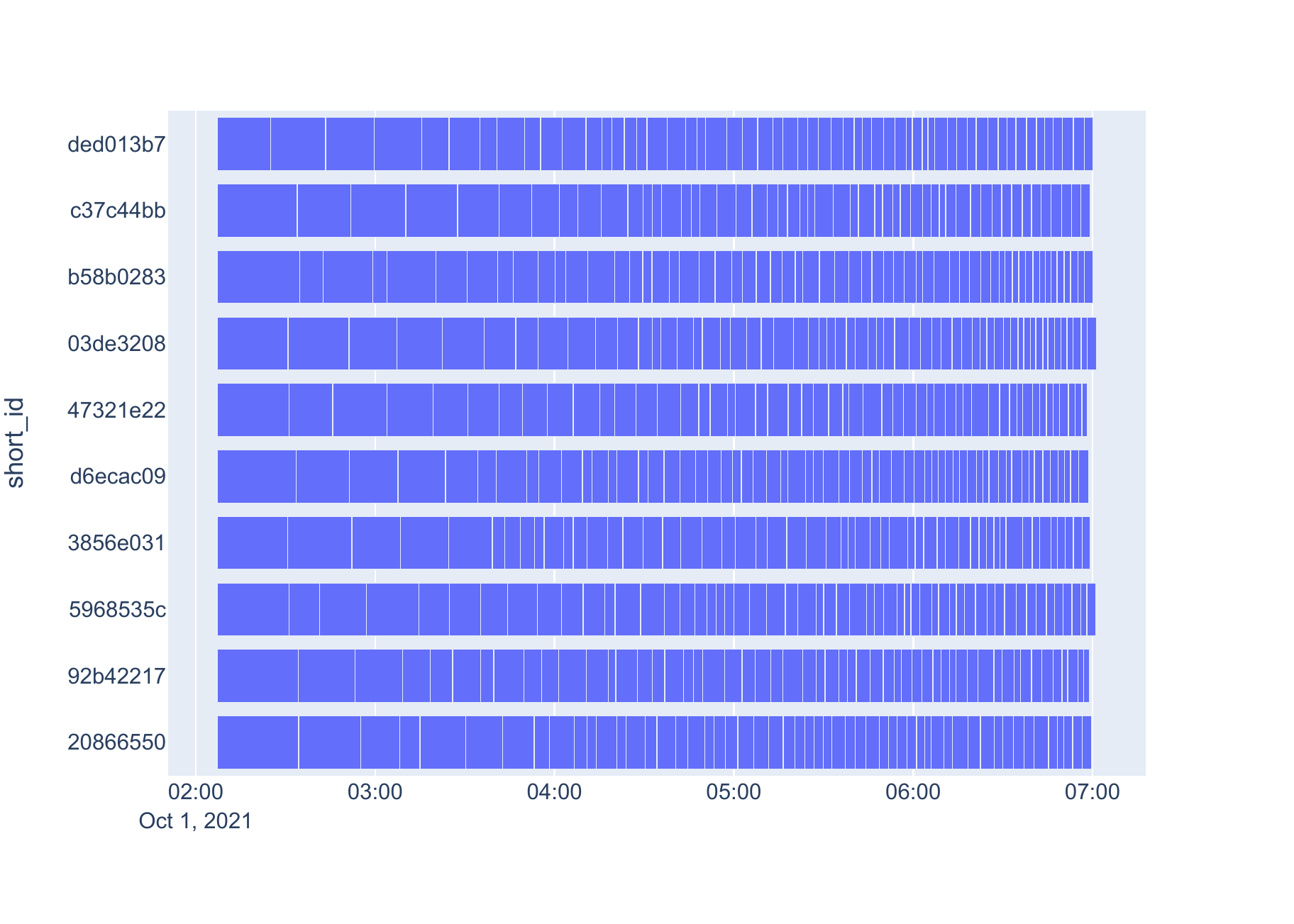}
    \caption{Distribution of AlphaFold processing among the Dask workers, where each row denotes the processing load for each worker, designated by a shortened portion of its UUID. Blue bars indicate time spent running AlphaFold; white dividing lines the Dask scheduler overhead. The results of 10 workers are shown out of 1200 and are representative of the whole.}
    \label{fig:workers}
\end{figure}

Fig.~\ref{fig:workers} shows how the inference processing workload was distributed among Dask workers over an approximately five hour run. Each row corresponds to a single Dask worker where a portion of their UUID is given as a ``short\_id'' to differentiate each worker and otherwise has no bearing on given worker's capabilities. The blue blocks denote the time spent processing an individual protein, and the white dividing lines indicate the overhead of switching between proteins. Note that only 10 randomly selected workers are shown from the full set of 1200 workers for the sake of brevity, but are representative of the observed behavior for all of them. The first set of proteins for each worker took significantly longer to process than those at the end due to task sorting by length as described in~\ref{sub:workflow}; given that all the Dask workers finished all of their respective tasks within minutes of one another indicates that sorting, together with the Dask dataflow execution model, provides effective load balancing and distribution, despite the heterogeneously sized inputs. Workflows using up to 1000 Summit nodes (6000 GPUs/Dask workers) were successfully deployed; to our knowledge this is one of the largest deployments of Dask on Summit to date.

\subsubsection{Proteome-scale structure prediction for \textit{S. divinum}}
Like other eukaryotic genomes, the protein sequences in the proteome of \textit{S. divinum} are more challenging to model than those of prokaryotic genomes. This proteome is unique in the sense that the vast majority of its sequences have not been publicly released on the either the National Center for Biotechnology Information (NCBI) database or Uniprot. Search in the RCSB PDB experimental structural database returns only 37 sequences with a sequence identity of 90\% or above. Therefore, our models can be considered virtually all new structures; while many of them would most likely be structurally similar to existing known folds, subtle but potentially functionally important differences have yet to be explored. In terms of model confidence/quality, using the pLDDT scores and considering the top-ranked models, about 57\% of sequences have a mean pLDDT score at greater than 70, which is considered high-quality. The total coverage of high-confidence pLDDT across all residues of the full proteome is 58\%, and about 36\% with an ultra-high confidence pLDDT $>$ 90. Using the global metric pTMS, about 53\% of top-ranked models have a pTMS greater than 0.6, a threshold indicating high quality. The mean number of recycles of these top-ranked models is 12. Total compute costs are about 2000 Andes node hours for the feature generation step, and 3000 Summit node hours for model inference, including the overheads.

\subsection{Geometry Optimization Method}
\begin{figure}[t]
    \centering
    \includegraphics[width=\linewidth]{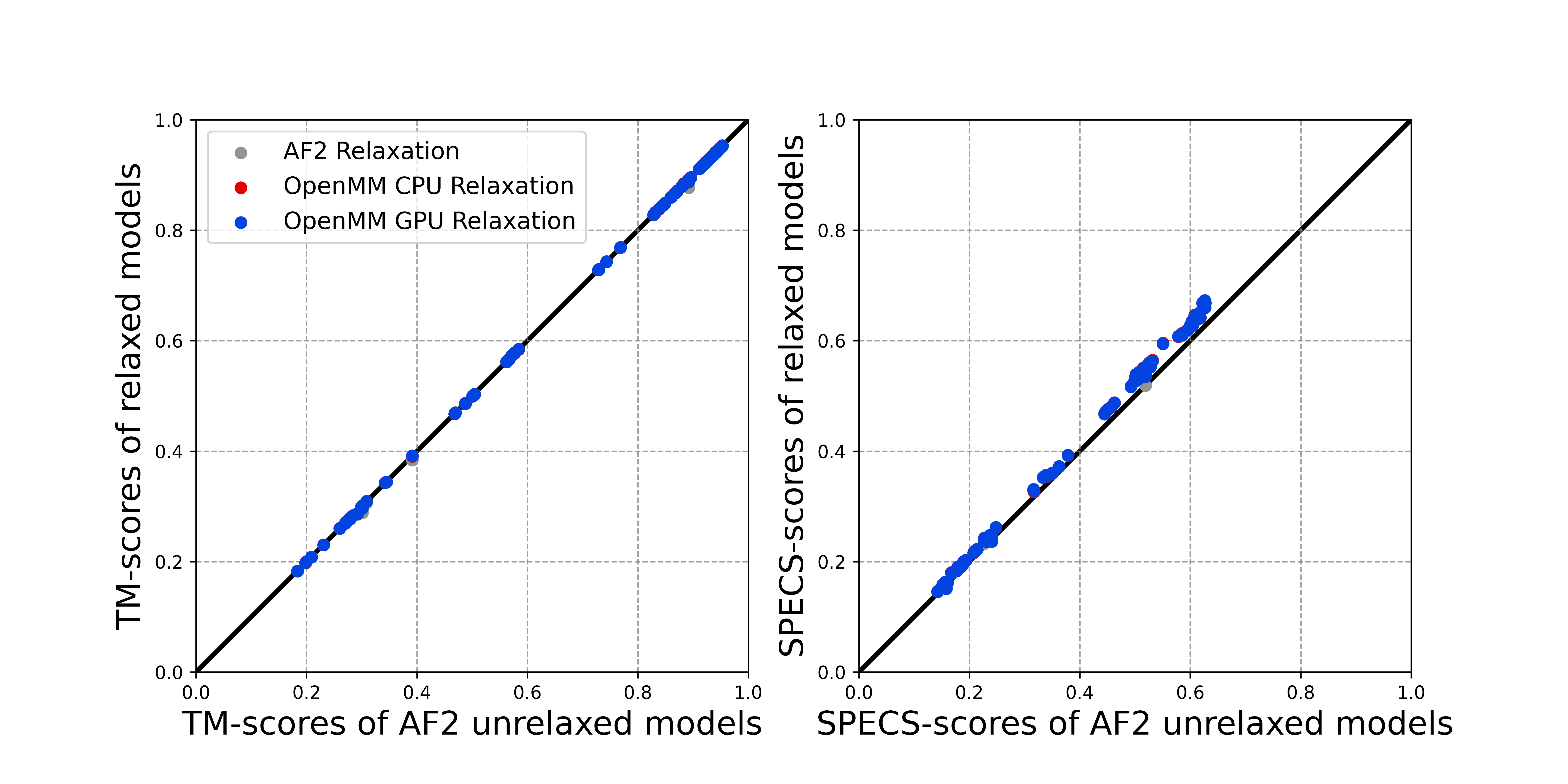}
    \vspace{-0.5cm}
    \caption{Correlation plots of structural metrics for unrelaxed models relaxed models using the three minimization methods, for (left) TM-score and (right) SPECS-score results. 
    } 
    \label{fig:specs-tm}
\end{figure}

The capabilities of the performance-optimized relaxation method reported here are quantified and compared to the AlphaFold2 results for a subset of CASP14 targets. Specifically, 19 of the CASP14 target sequences have publicly available high resolution crystal structures. For each model associated with this subset of CASP14 targets, we have an unrelaxed model saved as an intermediate output from the AlphaFold2 run, the AlphaFold2 relaxed model, and relaxed models produced from the optimized OpenMM method using CPUs and GPUs. The template modeling score (TM-score) \cite{zhan2004tmscore} and superposition-based protein embedded C$\alpha$-sidechain score (SPECS-score) \cite{Alapati2020} are used to quantify the quality of model relative to the experimentally determined protein structure. Fig.~\ref{fig:specs-tm} shows the TM-score (left) and SPECS-score (right) of relaxed models versus the respective scores of the AlphaFold unrelaxed models. TM-score addresses only backbone atoms, while SPECS-score reports on sidechain position also. Based on the strong correlation between structural metric values of the unrelaxed and relaxed models, we can see that no major structural changes have been made in the clash-removal procedure, as is required to preserve the inferred structure; importantly, no decreases in these metrics are seen. However, SPECS-scores do improve slightly after minimization for structural models that already have high SPECS-scores, indicating that sidechain positions are being further optimized towards their native crystallographic values by the molecular mechanics force field. All three relaxation methods recover equivalent model quality, which indicates that the additional steps used in the original AlphaFold relaxation method do not ensure higher quality models and, so, are not necessary. 

Beyond the TM-scores and SPECS-scores, we also quantified the reduction in structural violations such as clashes and bumps defined in \ref{sub:geom} after a relaxation method is applied. The full set of CASP14 targets (160 models in total) were analyzed to gather the average reduction in both types of violations. The AlphaFold unrelaxed models had, on average, 0.22 $\pm$ 1.09 clashes with a maximum of 8 clashes in a single structure. For all three minimization methods, clash violations are completely removed; this is unsurprising since C$\alpha$-C$\alpha$ atomic distances less than 1.9 \AA\ are energetically unfavorable within the molecular mechanics force field and are quickly resolved by the energy minimization procedure. For the smaller bump violations, the unrelaxed models had an average 3.76 $\pm$ 12.74 bumps with an observed maximum of 148 in a single structure. Application of the AlphaFold2, OpenMM GPU, and OpenMM CPU relaxation methods resulted in a reduction of bumps down to an average of 2.12 $\pm$ 3.70 (max of 26), 2.71 $\pm$ 5.90 (max of 58), and 2.59 $\pm$ 5.34 (max of 8) bumps, showing that relaxation also consistently reduces bumps. It should be noted that the optimization algorithm that underlies all three methods is not deterministic; variability in violation counts between different runs with the same method is expected. \vspace{-0.5em}

\begin{figure}[htbp]
    \centering
    \includegraphics[width=\linewidth]{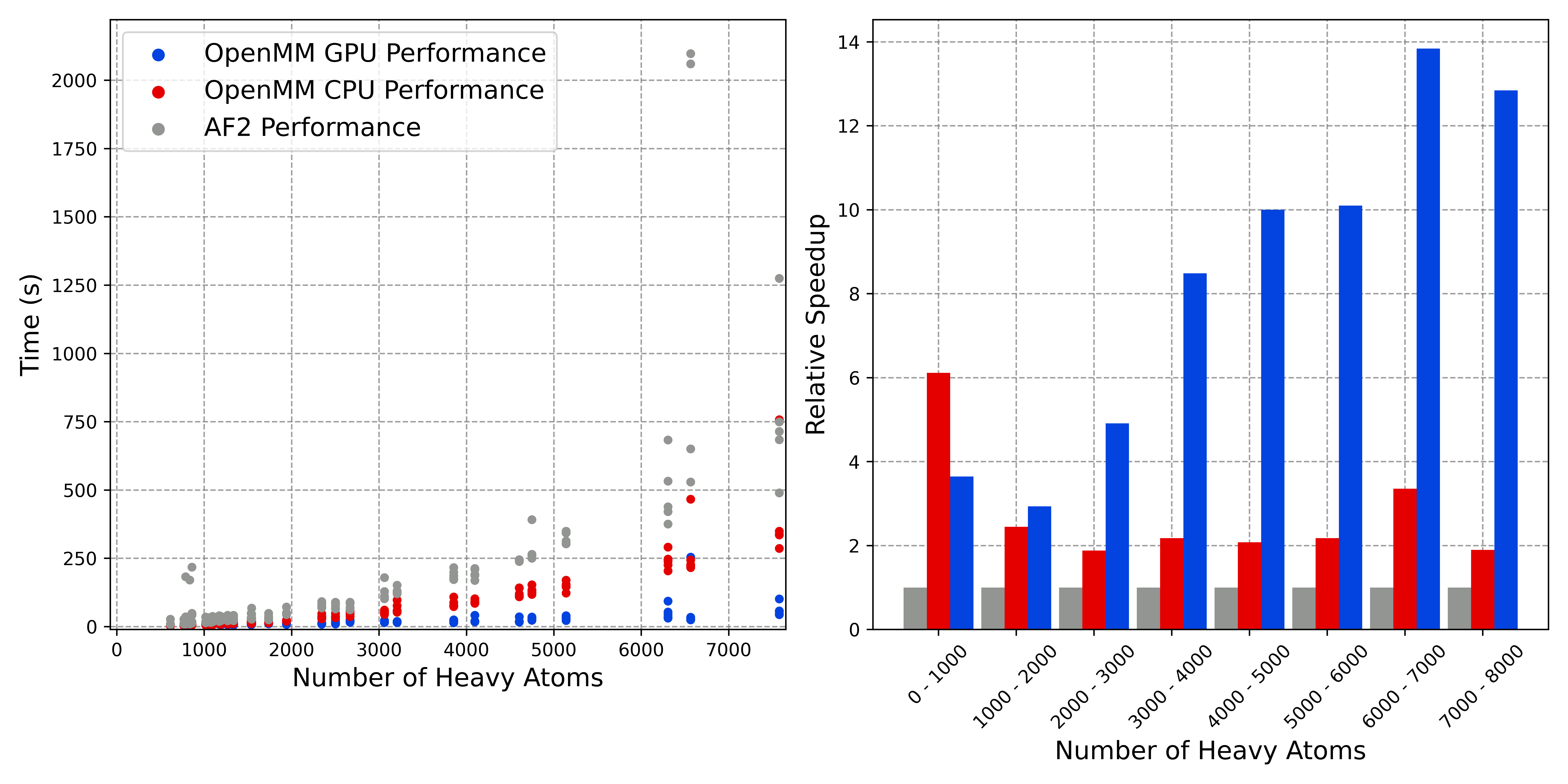}
    \caption{Computational efficiency for relaxation methods.
    (A) Time to solution results for relaxation of AlphaFold2 (AF2) predicted CASP14 targets with the AF2 relaxation method (grey) and  with our method run on OLCF Andes CPU nodes (red) and on OLCF Summit GPU nodes (blue). 
    A large outlier in the AF2 data is not included in timing results, where a relaxation calculation of T1080 took close to 4.5 hours. 
    (B) Computational speedups for these jobs, relative to the average timing for the AF2 method, across a range of system sizes. 
    } 
    \label{fig:min_comp_eff}
\end{figure}

\subsection{Geometry Optimization Workflow}
The CASP14 results indicate that our methodology, optimized for the high-throughput HPC setting by removal of extraneous quantification of violations and the potential for multiple, unnecessary iterations of calculations, and use of GPUs, results in equal success in reduction of crystallographic violations. Here we show the performance gains realized with this method. 

Fig. ~\ref{fig:min_comp_eff} depicts the time-to-completion results for the CASP14 targets in relation to the systems' total number of heavy (non-hydrogen) atoms, which is a better metric to quantify size of a job in a molecular mechanics calculation than the number of residues. As the number of heavy atoms increases, so does computational cost; however our optimized OpenMM implementation run on the GPUs was able to achieve marked performance boosts relative to the original AlphaFold relaxation method. Execution of the minimization calculations using a Dask workflow on Summit's GPU nodes provides up to a 14-fold speedup. Notably, a single structure (T1080) took close to 4.5 hours using the original AlphaFold method. This degree of speed up indicates that one of the limiting bottleneck steps from the original AlphaFold workflow has been removed, enabling the application of this protein structure prediction workflow to the proteome-scale. To further demonstrate the performance of the optimized relaxation method, geometry optimization calculations were performed on the genome-scale set of top structural models (3205) for 
\textit{D. vulgaris} Hildenborough. Relaxation of the 3205 \textit{D. vulgaris} Hildenborough structures was completed in 22.89 minutes using 8 Summit nodes with 6 Dask workers per node (48 workers in total). 

\subsection{Proteome-scale data analysis}
Our research on organisms important to DOE research includes a strong focus on the determination of the functions of proteins with no known function. These ``hypothetical" proteins are usually those with a sequence identity match below 20\% to annotated proteins in databases \cite{gao2021high}. For those with match below 10\%, even the state-of-the-art HMM-based methods fail. The use of a predicted structure to find homology to annotated proteins is now becoming a viable option, as the structures of proteins are often more conserved than their sequences. To demonstrate this type of analysis, we performed a structural TM-score based alignment against the pdb70 database using the global alignment module of the APoc program \cite{gao2013apoc} on the set of 559 proteins labeled as ``hypothetical" in the March 2020 NCBI GenBank proteome file for \textit{D. vulgaris} Hildenborough. A top alignment TM-score of 0.60 or above was found for 239 predicted structures, 215 of which had a sequence identity match of $<$ 20\%, and 112 for match $<$ 10\%, with the majority of aligned proteins having useful annotations. This suggests that a wealth of potential new information providing clues to the functions of these unknown proteins could be provided from matching high-confidence predicted structures to existing crystallographic databases.

An additional important use for this proteome-scale approach is to find potential new folds (tertiary structure) or novel quaternary structures in multi-domain proteins, and possibly, help discover new metabolic pathways. For example, on the same set of 559 hypothetical proteins, there were several instances of predicted structures with very high model confidence scores and very poor TM alignment scores to pdb70. One of these, with over 98\% of residues having pLDDT scores of over 90, has a top TM-score of 0.358 against the pdb70. Further analysis of this sequence revealed that it, and its homologs, had been independently annotated by experimental methods which discovered a novel mechanism for homocysteine synthesis \cite{allen2015homocysteine,price2018filling}. Therefore, it is possible that high-confidence predicted structures with no strong structural matches to any experimental structures may provide leads for the discovery of new enzymatic pathways. 

\section{Implications and Conclusion}
\label{sec:discussion}
Accurate structure predictions for all proteins in an organism's proteome, and collection of these proteome-scale models into large databases across species, has profound implications for biological research, as recognized by the AlphaFold Database project \cite{alphafold2021DB}. Here, we have demonstrated that the resources of the DOE leadership computing facilities can be harnessed to help in this effort. In fact, with our optimized pipeline for Summit and Andes, we have the ability to predict the structure of over 25,000 proteins in a plant proteome in a matter of hours, including all three stages of the pipeline from input generation to the final geometry optimization. Our optimizations for high-throughput deployment of AlphaFold on Summit were also included in AF2Complex \cite{gao2021predicting}, which is a generalization of AlphaFold that extends the model inference to prediction of protein-protein complexes, using the DL models trained for predicting either single protein sequences or complexes. The prediction of accurate protein complex structures at scale is an exciting new possibility especially relevant to HPC computing due to a quadratic (or higher) order dependence on the number of protein sequences.

It should be mentioned that while the CPU-based feature-generation step required fewer total node hours than the model inference step, the total wall times were higher, due to the fact that Andes, an analysis cluster, does not contain as many nodes as Summit and that the queue policies for Andes favor small, long jobs rather than large, shorter jobs as is the case on Summit. Rather than marking the need for larger CPU clusters with different queue policies, however, the problem begs the attention of the community for considering the use of GPU acceleration for MSA programs. In fact, GPU implementations of HMMER were first reported over a decade ago with one version reported in 2009 achieving a 38-fold speedup over the native CPU HMMER \cite{walters2009evaluating}. Unfortunately, none of these implementations seem to have been seriously considered for adoption by the developers of the two most commonly used programs for this calculation, HMMER and HHSuite. 

We have also demonstrated the as-yet unexplored possibilities that these large structural datasets can provide for understanding protein function and structure-function relationships. In conclusion, we anticipate a paradigm shift in biological research with the recent accuracy of DL structure prediction coupled with HPC workflows, beginning a new era of big data and large-scale computing in biology.

\section*{Acknowledgments}
This research was sponsored in part by the Office of Biological and Environmental Research's Genomic Science program within the US Department of Energy Office of Science, under award number ERKP917, the Laboratory Directed Research and Development Program at Oak Ridge National Laboratory (ORNL), and used resources of the Oak Ridge Leadership Computing Facility, which is a DOE Office of Science User Facility supported under Contract DE-AC05-00OR22725, granted in part by the Advanced Scientific Computing Research (ASCR) Leadership Computing Challenge (ALCC) program, resources supported by the Partnership for an Advanced Computing Environment (PACE) at Georgia Tech. We thank Bryan Piatkowski, Jerry Parks and Justin North for genome information.

\bibliographystyle{unsrt}  
\bibliography{ada,coletti,software,hardware}

\end{document}